**Polarization–dependent resonant phenomena in all–dielectric scatterers: inversion of magnetic inductance and electric displacement**

**Aleksandr Shvartsburg[1, 2] and Sergey Artekha[2, a]**

[1] *Joint Institute for High Temperatures of RAS, Moscow 125412, Russia*

[2] *Space Research Institute of RAS, Profsoyuznaya 84/32, Moscow 117997, Russia*

[a] e-mail: sergey.arteha@gmail.com  (corresponding author)

**Abstract**

The theoretical description and experimental verification of resonant phenomena in electromagnetic fields generated in the near zone of all-dielectric rectangular thin sub wavelength frames, subjected to an incident microwave, are considered. The geometry of considered problems is presented by means of arrangements of these frames in three orthogonal planes, normal respectively to electric component, magnetic component and wave vector of the incident wave. Such trio paves the way to design of 3D all-dielectric multiresonant microwave unit cell. Displacement currents, generated by linearly polarized electromagnetic waves in the system, lead to the formation of magnetic and electric dipoles, each of which possess their own resonant frequencies. Resonant inversion of magnetic inductance and electric displacement is observed. The sliding incidence of plane wave on the frame is shown to provide the sharp and deep resonance in the components of generated field. The phase shift equal to $\pi$ between the magnetic components of incident and generated wave indicates the formation of negative magnetic response. There is observed the angular anisotropy of arising dipoles, manifested in values of resonance frequencies and in the dependence of depths of resonant spectral dips upon the orientation of dipoles, with respect to the direction of propagation of incident wave.

**Some terms used (for the second version on arXiv.org):**

**Displacement current density** is the quantity $\partial \mathbf{D}/\partial t$ appearing in Maxwell's equations that is defined in terms of the rate of change of **D**, the electric displacement field.

**Inductance** (or self-induction coefficient) is the coefficient of proportionality between the electric current flowing in any closed circuit and the total magnetic flux created by this current through the surface bounded by this circuit.

**Mutual induction** is the phenomenon of the occurrence of an induction EMF in one circuit when the current strength in the second circuit changes (conductors are said to be inductively coupled or magnetically coupled).

**Inversion** is a change in the direction or sign of a physical quantity or effect to the opposite.

**Mie scattering** is the Mie solution to Maxwell's equations describes the scattering of an electromagnetic plane wave by a homogeneous sphere. The solution takes the form of an infinite series of spherical multipole partial waves.

**A metamaterial** is a type of material engineered to have a property, that is derived not from the properties of the base materials but from their newly designed structures.

**Subwavelength** means sizes $r \leq \lambda$.

**The near zone** is the region within sizes r ≪ λ, and the far zone is the region for which r ≫ 2λ. The near field and far field are regions of the electromagnetic (EM) field around an object. Non-radiative near-field behaviors dominate close to the object, while electromagnetic radiation far-field behaviors predominate at greater distances.

**The optical range** (visible spectrum) is generally defined to encompass electromagnetic radiation with wavelengths in the range from 380 nm to 750 nm, or frequencies in the range from 400 THz to 670 THz.

**Infrared range** includes electromagnetic radiation with wavelengths λ between 750 nm and 1 mm, which corresponds to a frequency range from 300 GHz to 400 THz.

**curl**, also known as **rotor**, is a vector operator in vector calculus that describes the infinitesimal circulation of a vector field.

**1. Introduction**

The last 10 – 15 years have witnessed a rapid growth of interest in the phenomena of resonant electromagnetic inductance for all–dielectric structures irradiated by electromagnetic waves. Variations of magnetic components of these waves were shown to induce the displacement currents in the dielectric material; these currents, in their turn, generated the magnetic modes. The efficiency of this generation proved to be strengthened drastically, when the frequencies of driving waves were approaching to the resonant frequencies of dielectric structures determined by their shapes, sizes and dielectric properties. Analysis of these phenomena constitutes nowadays one of the newly shaping branches in the electrodynamics of continuous media [1]. Although the dielectric ring resonators supporting the concentric GHz modes were considered several decades ago [2], the systematic research of magnetic modes became a "hot" topic later due to discovery of "artificial magnetism" [3, 4]. This unusual phenomenon was demonstrated by resonant excitation of magnetic modes and formation of negative magnetic inductance in all–dielectric sub wavelength rings in the field of monochromatic electromagnetic wave [5, 6]. The resonant frequencies were calculated for isotropic spheres [7] and rotationally symmetric particles [8]. The elliptically shaped plane circuit, providing the angular anisotropy of magnetic mode, was examined [9]. These facilities can be viewed as the generalizations of well-known Thomson oscillating circuit [10], combining the inductance and capacitive properties in one resonant *LC* structure. An another group of optical magnetization effects connected with Mie resonant scattering of electromagnetic waves [11] was considered for the dielectric discs [12], cylinders [13, 14] and spheres [15, 16]. Consideration of all–dielectric magnetic and electric dipoles in the microwave range is linked usually with the design of frame aerials and their radiation patterns in the far zone.

To the contrary, this paper aims at the theoretical analysis and experimental verification of the lowest *LC* resonances, both electrical and magnetic, induced in the thin all–dielectric sub-wavelength rectangular circuits by the driving linearly polarized GHz microwaves. 3D-structures with predefined properties can be constructed from such circuits (see Fig. 1).

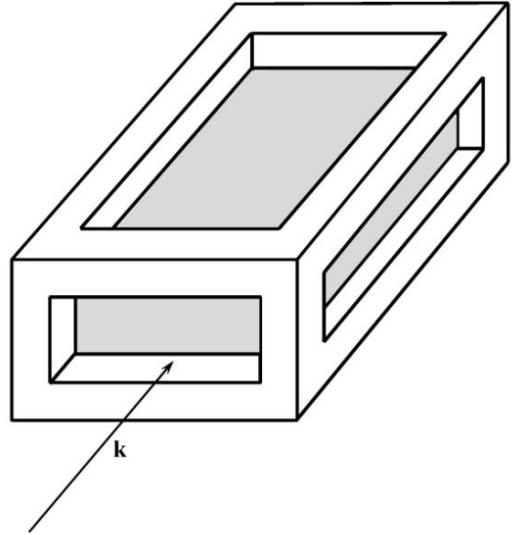

**Fig. 1.** 3D all-dielectric multiresonant microwave unit cell. Rectangular all–dielectric circuits are arranged in three orthogonal planes, normal respectively to electric component $\vec{E}$, magnetic component $\vec{H}$ and wave vector $\vec{k}$ of the driving wave.

The geometry of this problem can be presented by means of arrangements of these circuits in three orthogonal planes, normal respectively to electric component $\vec{E}$, magnetic component $\vec{H}$ and wave vector $\vec{k}$ of the driving wave. This trio paves the way to design of 3D all-dielectric multiresonant microwave unit cell, characterized by simultaneous formation of magnetic and electric resonant dipoles, admitting the inversion of magnetic inductance and electric displacement. We'll examine the resonances in rectangular circuits arranged in two different positions A and B in each of these three planes (see Fig. 2) and, thus, distinguished by their orientation with respect to the vectors $\vec{E}$, $\vec{H}$, $\vec{k}$. The rectangular frame can be oriented towards the vector of interest to us either by its larger side or by its smaller side.

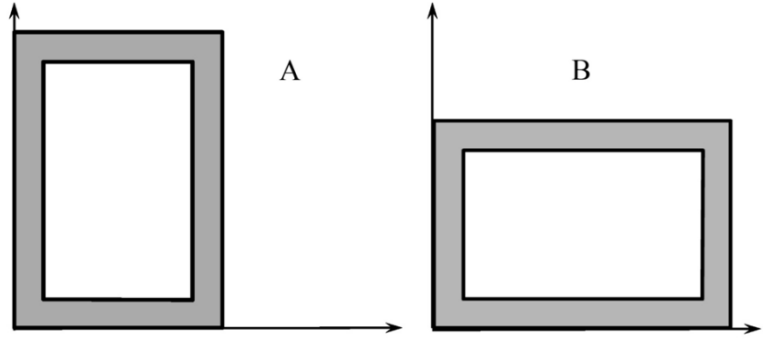

**Fig. 2.** Positions A and B of rectangular frames in each of three mutually orthogonal planes. The rectangular frame can be oriented towards the vector of interest to us either by its larger side or by its smaller side.

The current interest to the electromagnetic inductance in dielectric frames in this geometry, viewed as a promising platform for 3D integrated microwave devices, is threefold:

**i.** controlled excitement of both electric and magnetic resonances in the different sides of rectangular frames in 3D microwave devices;

**ii.** feasibility of creation of anisotropic radiation patterns in different spectral ranges;

**iii.** elaboration of miniaturized sub wavelength circuits sensitive to the directivity and polarization of incident radiation [17-19].

Several effects contribute to the response of the system to an external irradiation [20]; each of them – at its own resonant frequency. To compare and contrast the different $LC$ resonances we consider the resonant phenomena for the rectangular frame with internal sizes $a$ and $b$ $(a \geq b)$ and the square cross-section size $h$. Figure 3 shows the location of the frame in the $(x, y)$ plane, i.e. normal to the magnetic component of incident wave. The sizes of an auxiliary rectangular contour formed by the axes of rods constituting the rectangular frame are $a_1 = a + 0.5h$ and $b_1 = b + 0.5h$; the external sizes of this frame are $a_2 = a_1 + 0.5h$ and $b_2 = b_1 + 0.5h$.

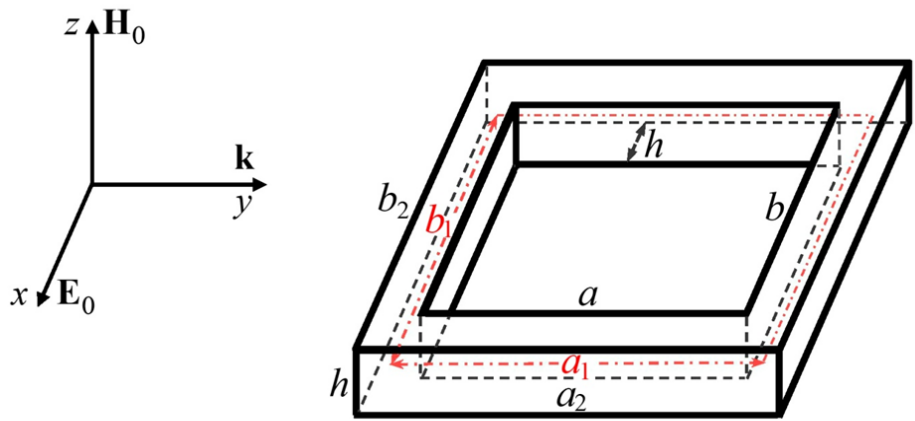

**Fig. 3.** Frame parameters and the geometry of scattering for sliding incidence: the plane of frame $(x, y)$ is normal to the magnetic component of incident wave $\mathbf{H}_0$. In this case, vectors $\mathbf{E}_0$ and $\mathbf{k}$ are located in the plane of the frame.

The model of thin circuit, corresponding to the smallness of cross-section size $h$ with respect to the distance between the rod's axes $d$ $(h \ll d)$, is used in the analysis below; herein $d = b_1$ and $d = a_1$ in positions A and B respectively (see Fig. 2). The value of dielectric permittivity $\varepsilon$ in the GHz range under discussion is taken as high as $\varepsilon = 150-160$. The paper is organized as follows: the resonant inversions of magnetic inductance and electric displacement in the rectangular frame subject to its orientation with respect to the incident microwave are considered theoretically in Sects. 2 and 3 respectively; the experiments verifying this analysis are briefly described in Sect. 4; the main results are listed in Sect. 5. Some general mathematical formulas are shown in the Appendix.

## 2. Magnetic resonances and negative magnetic inductance in the rectangular all–dielectric frames

Analysis of magnetic resonances is carried out below for two cases characterizing 3D orientation of rectangular frame: the plane of frame $(x, y)$ is either normal (Fig. 3) or parallel (see Fig. 4 and Fig. 5) to the magnetic component $\vec{H}$ of incident wave ($z$ – direction).

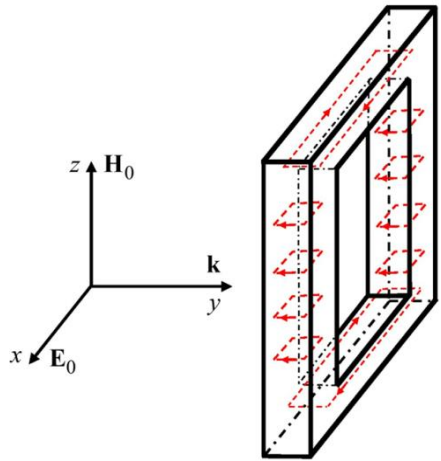

**Fig. 4.** The geometry of scattering for normal incidence: the plane of frame $(x, z)$ is parallel to the magnetic component of incident wave. Unlike the sliding incidence shown in Figure 3, here the wave $\mathbf{k}$ falls perpendicular to the frame. In this case, the fields $\mathbf{E}_0$ and $\mathbf{H}_0$ lie in the plane of the frame.

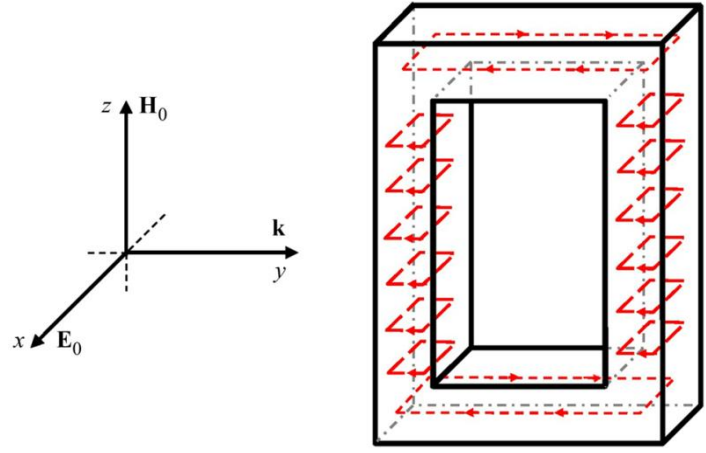

**Fig. 5.** The geometry of scattering for sliding incidence **k**: unlike the case in Fig. 3, the plane of frame $(y,z)$ is parallel to the magnetic component $\mathbf{H}_0$ of incident wave, but the field $\mathbf{E}_0$ is perpendicular to the frame.

**2a.** To examine the magnetic response of all-dielectric rectangular frame induced by the sliding incidence of microwave one has to consider the alternating magnetic flow $\Phi$, stipulated by the auxiliary frame self-inductance (see Fig. 3). This flow is known to generate in the frame the electromotive force $U$ and the curl of an electric field $E_i$, directed along the frame's perimeter

$$U = -\frac{1}{c}\frac{\partial \Phi}{\partial t}, \quad E_i = -\frac{1}{2(a_1 + b_1)}\frac{\partial \Phi}{\partial t}. \tag{1}$$

This field $E_i$ generated in the circuit with dielectric permittivity $\varepsilon$, forms the alternating electric displacement $D = \varepsilon E_i$, providing in its turn the curl of a displacement current $I_d$ in the frame [2]

$$I_d = \frac{S_0}{4\pi}\frac{\partial D}{\partial t}, \tag{2}$$

here $S_0 = h^2$ is the area of square cross-section of the frame's sides. Substitution of $D$ with Eq. (1) into Eq. (2) brings the value of displacement current in the frame

$$I_d = -\frac{h^2 \varepsilon}{8\pi c(a_1 + b_1)}\frac{\partial^2 \Phi}{\partial t^2}. \tag{3}$$

The magnetic flow $\Phi_i$, induced by the displacement current $I_d$ from Eq. (3), can be expressed by means of self-inductance of this frame $L$:

$$\Phi_i = \frac{L I_d}{c}. \tag{4}$$

To find the unknown magnetic flow $\Phi$, one has to add the inductance flow $\Phi_i$ (4) generated by displacement current to the initial magnetic flow $\Phi_0(t) = H(t) S_{frame}$ formed by the magnetic component of incident wave $H = H_0 \exp[i(ky - \omega t)]$. Considering, e.g., the orientation of frame, corresponding to the case A (see Figs. 2 and 3), we'll find

$$\Phi_0 = b_1 \int_{-a_1/2}^{a_1/2} H dy = H_0 a_1 b_1 F_a \exp(-i\omega t), \text{ where } F_a = \frac{\sin(ka_1/2)}{(ka_1/2)}. \tag{5}$$

It follows from (3) and (4) that

$$\Phi_i = -\frac{h^2 \varepsilon L}{8\pi c^2 (a_1 + b_1)} \frac{\partial^2 \Phi}{\partial t^2}.$$

We substitute $\Phi_i = \Phi - \Phi_0$ into the last expression, introduce the definition of $\omega_0$ and express the second derivative from the resulting equality. Such manipulations bring the equation governing the unknown flow $\Phi$:

$$\frac{\partial^2 \Phi}{\partial t^2} + \omega_0^2 \Phi = \omega_0^2 \Phi_0, \text{ where } \omega_0^2 = \frac{2\pi c^2 (1+p)}{h^2 \varepsilon K}. \tag{6}$$

Dimensionless factor $K$ in Eq. (6) determining the self-inductance of rectangular frame $L = 4a_1 K$ subject to the ratio of distances between the axes of its sides $p = b_1/a_1$ is given in Appendix (A4). Solution of Eq. (6) reads as

$$\Phi = \Phi_0 \Lambda(\omega), \quad \Lambda(\omega) = \frac{\omega_0^2}{\omega_0^2 - \omega^2}. \tag{7}$$

Equation (7) describes the forced oscillations of magnetic flow $\Phi$ passing through the circuit, $\omega_0$ is the resonant frequency $\omega_0 = 2\pi f_0$, $\Lambda(\omega)$ is the resonant factor.

The structure of the fields that arise as a result of this effect is shown in the dimensionless form for the components $E_x$, $E_y$, $H_z$, respectively, in Fig. 6, Fig. 7 and Fig. 8.

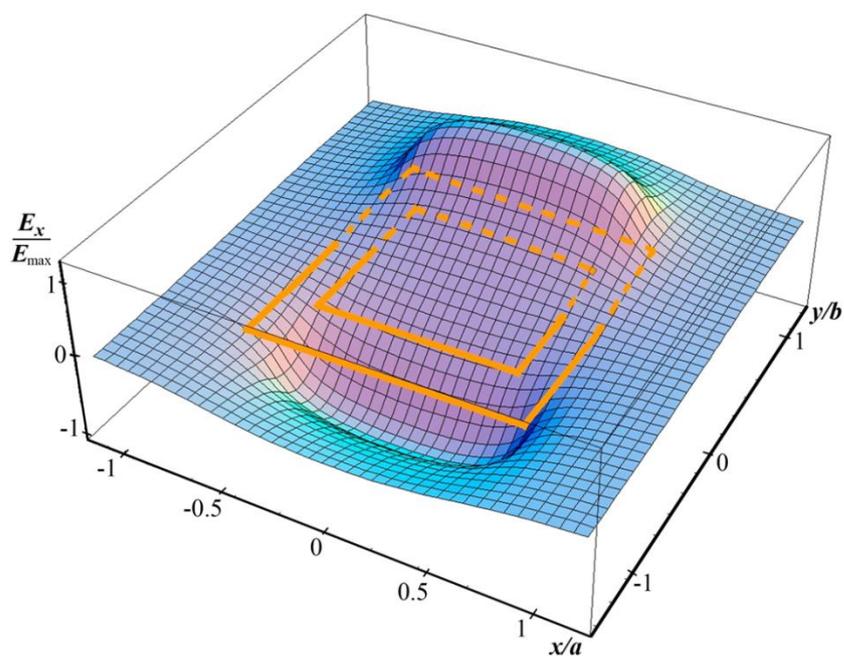

**Fig. 6.** Distribution of the component $E_x$ in the plane of the frame, which is conventionally drawn by a solid line at the zero level of the field component.

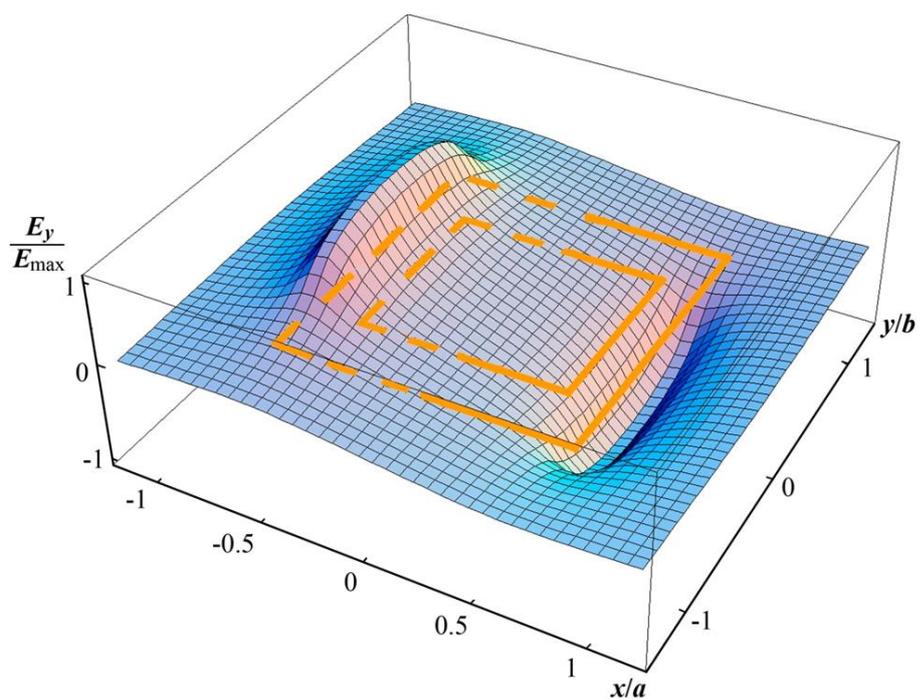

**Fig. 7.** Distribution of the component $E_y$ in the plane of the frame, which is conventionally drawn by a solid line at the zero level of the field component.

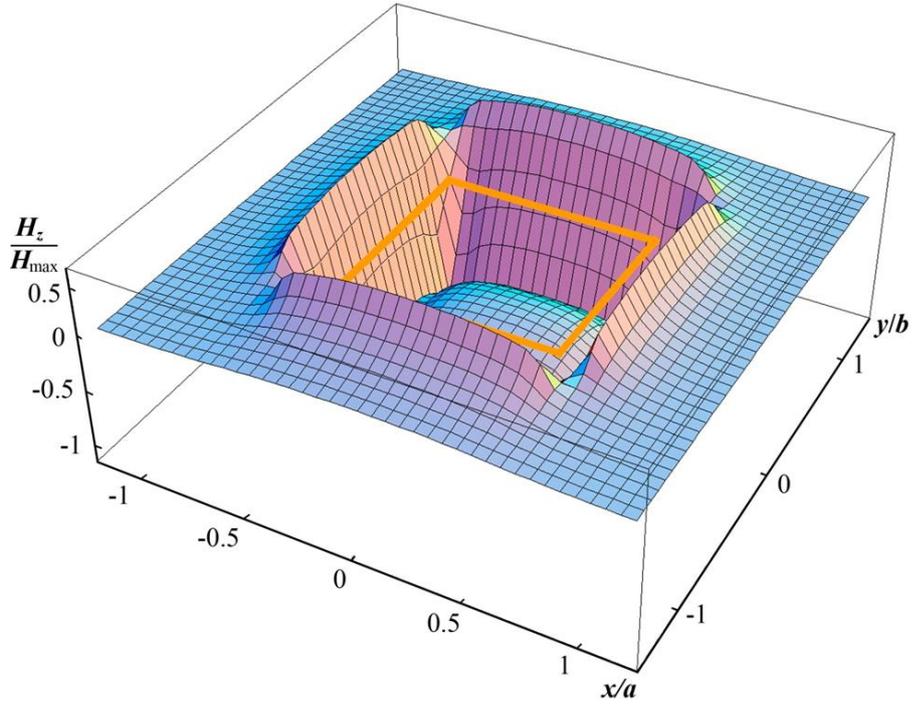

**Fig. 8.** Distribution of the component $H_z$ in the plane of the frame, which is conventionally drawn by a solid line at the zero level of the field component.

Note, that Eqs. (5) and (6) are valid for the orientation B (see Fig. 2) due to replacement $a_1 \Leftrightarrow b_1$, $F_a \to F_b$, herein the resonant frequency $f_0$, determined in (6), remains unchanged. The calculated frequency $f_0 = 1.75$ GHz is in good agreement with the experimental data.

**2b**. Consider now the electromagnetic inductance in a rectangular frame, when the alternating magnetic component $\vec{H}$ of incident wave is parallel to one of the rods, forming the frame. In such geometry, shown in Figs. 4 and 5, the field $\vec{H}$, directed along z-axes, generates the curl of electric currents and fields $\vec{E}_\perp (E_x, E_y)$ distributed in the rod's cross-section ($E_z = 0$). The field $\vec{E}_\perp$, in its turn, induces the secondary magnetic field $\vec{H}_1(0, 0, H_{1z})$ in the rod itself. The similar inductive effects ensuring the generation of secondary magnetic field $H_2$ arise in the parallel rod too. This structure resembles the pair of long inductance coils with concentric currents and axial magnetic fields. However, in the framework of thin circuit model the magnetic interaction between these coils is small: $\left|\vec{H}_{12}\right|\left|\vec{H}_1\right|^{-1} \propto hd^{-1} \ll 1$. Neglecting this mutual influence, one can consider the fields in each rod $\vec{H}$, $\vec{H}_1$ and $\vec{E}_\perp$ linked by Maxwell equation

$$\text{rot}\,\vec{E}_\perp = -\frac{1}{c}\frac{\partial}{\partial t}\left(\vec{H} + \vec{H}_1\right). \tag{8}$$

Due to smallness of sizes of square cross–section of dielectric rod $h$ with respect to the wavelength of incident wave one can view the incident field $H$ inside the rod as spatially homogeneous: $H = H_0 \exp(-i\omega t)$. The fields $\vec{H}_1$ and $\vec{E}_\perp$ are linked by another Maxwell equation:

$$\operatorname{rot}\vec{H}_1 = \frac{\varepsilon}{c}\frac{\partial \vec{E}_\perp}{\partial t}. \qquad (9)$$

Considering the long rod and ignoring the end effects we obtain from Eqs. (8) and (9) the equation governing $H_{1z}$ component of induced magnetic field inside the rod

$$\frac{\partial^2 H_{1z}}{\partial x^2} + \frac{\partial^2 H_{1z}}{\partial y^2} - \frac{\varepsilon}{c^2}\frac{\partial^2 H_{1z}}{\partial t^2} = -\frac{\varepsilon\omega^2}{c^2} H_0 \exp(-i\omega t). \qquad (10)$$

Note, that the directions of induced field components $H_{1z}$ inside and outside of the rod are opposite. Thus, supposing the value $H_{1z}(x, y) = 0$ at each flank side of the rod $(-h_1 \leq x \leq h_1, -h_1 \leq y \leq h_1, h_1 = 0.5h)$ one can search the solution of Eq. (10) by means of Fourier series

$$H_{1z} = \exp(-i\omega t)\sum_m \sum_n \aleph_{mn} \cos\left(\frac{m\pi x}{2h_1}\right)\cos\left(\frac{n\pi y}{2h_1}\right), \quad m = 1;3;5..., \quad n = 1;3;5.... \qquad (11)$$

To find the coefficients $\aleph_{mn}$ it is expedient to present the incident field $H_0 \exp(-i\omega t)$ in a form similar to (11): the coefficient $\aleph_{11}$ related to the lowest mode in the presentation (11), is $\aleph_{11} = H_0 (4/\pi)^2$. The coefficients $\aleph_{mn}$ can be calculated due to orthogonality of terms in the Fourier series. Thus, the solution of Eq. (10), describing the lowest mode in the presentation (11) in the spectrum of oscillations of induced magnetic field, reads as:

$$H_{1z} = H_0 \left(\frac{4}{\pi}\right)^2 \Lambda(\omega)\cos\left(\frac{\pi x}{h}\right)\cos\left(\frac{\pi y}{h}\right)\exp(-i\omega t), \quad \Lambda(\omega) = \frac{\omega^2}{\omega_0^2 - \omega^2}. \qquad (12)$$

Here $\Lambda(\omega)$ is the resonant factor, $\omega_0 = 2\pi f_0$ is the lowest resonant frequency

$$f_0 = \frac{c}{h\sqrt{2\varepsilon}}. \qquad (13)$$

Substitution of expression for $H_{1z}$ from (12) into Eq. (9) brings the values of electric components $E_x$ and $E_y$. Note, that the induced fields $\vec{H}_1$ and $\vec{E}_\perp$ obey to the pair of Maxwell equations (8) – (9). Introducing the electric displacement in dielectric rod $\vec{D}_\perp = \varepsilon\vec{E}_\perp$, one can see that these fields obey to the another pair of Maxwell equations too:

$$\operatorname{div}\vec{H}_1 = 0, \quad \operatorname{div}\vec{D}_\perp = 0. \qquad (14)$$

The calculated frequency $f_0 = 4.19\,\text{GHz}$ is in good agreement with the experimental data. Note, that the values of $f_0$ (13) obtained for the model of a long dielectric rod, ignoring the end effects, prove to be independent upon the rod's length for both positions A and B.

Having studied magnetic resonances, we will now move on to the analysis of electrical resonances.

**3. Resonant pair of electric dipoles and negative electric inductance in the rectangular all–dielectric frames**

Let us examine the electric $LC$ resonances excited in the pair of parallel sides of rectangular frame directed along the electric field of incident microwave $\vec{E}$, as is shown in Figs. 3 and 4. The alternating field $\vec{E}$ induces the periodically oscillating positive and negative charges $\pm Q$ on the opposite ends of these rods as well as the electric fields $E\varepsilon^{-1}\exp[i(q_x x - \omega t)]$ and displacement currents propagating along the rods ($x$-direction). These currents can be presented as the waves with frequency $\omega$ and wave number $q_x$. To analyze these waves, it is convenient to model each rod as a segment of transmission line characterized by the capacitance $c_a$, self-inductance $l_a$ and mutual inductance $m_a$ per unit length. Unlike the insignificant mutual inductance of curl of currents in the parallel coil-like structures, discussed in the Sect. 2b, the magnetic interaction of linearly polarized displacement currents in the parallel rods is shown below to be an important effect, affecting the resonant regimes in the rectangular dielectric frame. Thus, we have to examine the pair of interacting electric dipoles. Considering, e.g., the position A and ignoring the losses, one can describe the voltage $U$ and displacement current $I_d$ in this segment with length $a_2 = a_1 + 0.5h$ (Figs. 3 and 4) and cross-section area $h^2$ by the equations of transmission line [10]:

$$\frac{\partial U}{\partial x} + \frac{l_a + m_a}{c^2}\frac{\partial I_d}{\partial t} = -\frac{E\exp[i(q_x x - \omega t)]}{\varepsilon}, \qquad (15)$$

$$\frac{\partial I_d}{\partial x} + c_a \frac{\partial U}{\partial t} = 0. \qquad (16)$$

The displacement current $I_d$ is linked with the induced electric displacement field $E_d$, directed along $x$-axis:

$$I_d = \frac{\varepsilon h^2}{4\pi}\frac{\partial E_d}{\partial t}. \qquad (17)$$

Substitution of solution for $I_d$ determined by set (15) – (16) to Eq. (17) brings the expression for $E_d$. Solution describing the standing half-waves of field $E_d$ in this rod near by the lowest resonance, corresponding to the value $q_x = \pi a_2^{-1}$, is

$$E_d = -\frac{4a_2^2 c_a \Lambda(\omega)}{\pi \varepsilon^2 h^2} E \exp[i(q_x x - \omega t)], \quad \Lambda(\omega) = \frac{\omega_0^2}{\omega_0^2 - \omega^2}, \quad \omega_0 = \frac{\pi c}{a_2 \sqrt{c_a (l_a + m_a)}}. \tag{18}$$

Polarization of the rod by field $E_d$ induces the oscillating charges $\pm Q$ on the ends of the rod:

$$Q = \frac{a_2^2 c_a \Lambda(\omega) E}{\varepsilon \pi^2} \exp(-i\omega t). \tag{19}$$

Here $\omega_0$ in (18) is the resonant frequency, $\Lambda(\omega)$ is the resonant factor. Thus, both the induced field $E_d$ and the polarity of electric dipole $P = Ql$ are inverted at the high frequency spectral ranges $f > f_0$.

Substitution of capacitance $c_a$, self-inductance $l_a$ and mutual inductance per unit length $\pm m_a$ from (A6) – (A8) into (18) determines the values $\omega_0 = 2\pi(f_0)_\pm$; the resonant frequencies $(f_0)_\pm$ relate respectively to the parallel and anti-parallel currents induced in the parallel rods

$$(f_0)_\pm = \frac{c}{h}\sqrt{\frac{\pi}{2\varepsilon[l_a \pm m_a]}}. \tag{20}$$

In a simple case, when the plane of frame is normal to the wave vector of incident wave $\vec{k}$ (see Fig. 4) the phases of currents excited in the parallel rods are equal; thus the currents are parallel and their mutual inductance is given by expression $+m_a$ from (A8). Calculation of amounts $l_a$ and $+m_a$ by means of formulas (A7) – (A8) for the rods with length $l = a_2$ and distance between their axes $d = b_1$ brings the resonant frequency $(f_0)_+^A = 4.09$ GHz. Analysis of resonances (20) in position B carried out by the same formulas (A7) – (A8) with the values $l = b_2$, $d = a_1$ gives the another resonant frequency $(f_0)_+^B = 4.72$ GHz. Thus, the LC resonant frequencies for the rectangular electric dipole, arranged in a plane, normal to the direction of wave propagation (see Fig. 4), are different in positions A and B.

To the contrary, analysis of resonances in the parallel rods in the geometry of Fig. 3 is more complicated, since the phase delays $\varphi$ between currents excited by the microwave with wavelength $\lambda$ in the parallel rods depend in this case upon the distances $d$ between them. Thus, the mutual inductance between rods becomes equal to $m_- = -m_a$ ($m_+ = +m_a$), when the distances between them are $d = 0.5\lambda$ ($d = \lambda$) and the phase delays constitute $\varphi = \pi$ ($\varphi = 2\pi$) respectively. Keeping in mind the sub wavelength frame let us restrict ourselves by the case $0 \leq d \leq \lambda$ $(0 \leq \varphi \leq 2\pi)$; herein the currents can change their directions during one period of incident wave $T$, providing the non-stationary piecewise variations of mutual inductances. These rapid variations impede the computation of resonant frequencies for the intermediate values of phase delay $0 < \varphi < 2\pi$.

To evaluate the influence of both inductances $m_-$ and $m_+$ on the resonant frequencies $(f_0)_\pm$ in all the range $0 \leq \varphi \leq 2\pi$ one can introduce the quantity $M$ presenting some value of mutual inductance averaged over the wave period $T$. This quantity $M$, dependent upon the frequency $f = T^{-1}$, in the cases $0 \leq d \leq 0.5\lambda$ $(M = M_1)$ and $0.5\lambda \leq d \leq \lambda$ $(M = M_2)$ is determined by formulae:

$$M_1 = m_-\left(\frac{2df}{c}\right) + m_+\left(1 - \frac{2df}{c}\right), \quad M_2 = m_-\left(2 - \frac{2df}{c}\right) + m_+\left(\frac{2df}{c} - 1\right). \tag{21}$$

The values of mutual inductances $M_1$ and $M_2$ calculated by means of Eq. (21) for the parallel rods, separated by distances $d = 0.5\lambda$ and $d = \lambda$ are $M_1 = M_2 = m_-$ and $M_1 = M_2 = m_+$. The intermediate values of $M$ (21) illustrating the frequency dispersion of mutual inductance subject to the phase factor $\varphi = 2\pi d \lambda^{-1}$ are shown in Fig. 9.

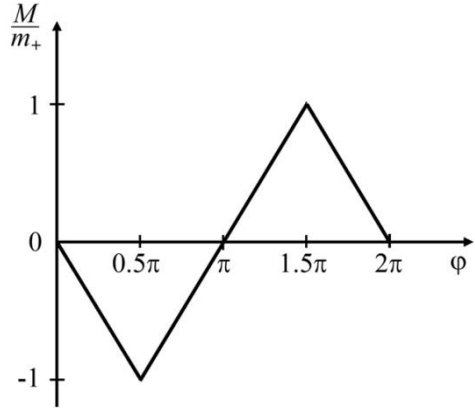

**Fig. 9.** Approximation of frequency dispersion of mutual inductance for the pair of interacting resonant electric dipoles.

Substitution of value $M(f)$ into Eq. (20) instead of $(m_a)_\pm$ brings the generalized equation governing the resonant frequencies $f_0$ for the wave – frame interaction under the conditions shown in Fig. 3. Thus, in case of position A $(d = b_1)$, this generalized equation reads as:

$$f_0\sqrt{l_a + M_a(f_0)} = \frac{c}{h}\sqrt{\frac{\pi}{2\varepsilon}}. \tag{22}$$

The resonant frequency of the pair of electric dipoles in position A computed from Eq. (22) is $(f_0)_A = 4.66$ GHz. Proceeding in the similar manner and using the values of inductances $m_-$ and $m_+$ in position B, we find the resonant frequency for position B: $(f_0)_B = 5.61\text{ GHz}$. Comparison with the experimental spectra shows the discrepancies of computed values of resonances with the experimental ones about 10-12% and 3-4% in positions A and B respectively. The greater discrepancy of computed frequencies $f_0$ in position A as compared with position B is linked with the fulfillment of condition $hd^{-1} \ll 1$, mandatory for the thin circuit model: since this condition is fulfilled better in position B $(ha_1^{-1} = 0.15)$, than in position A $(hb_1^{-1} = 0.25)$, the results predicted by this model for position B are more correct than for position A.

After the theoretical analysis of magnetic and electrical resonances, we move on to a brief discussion of the results from the viewpoint of experimental verification.

## 4. Brief description of experiments

Currently, when creating metamaterials with a negative magnetic response, metal elements of nanoscale values are used, the expected characteristics of which are checked in advance on a mathematical model to determine the integral field of interaction with the incident wave. Three-dimensional periodic structures with a period of hundreds of nanometers are built from such elements. However, the quality factor of such elements is limited by ohmic losses in the metal parts. With a large dielectric constant and low dielectric losses, the dimensions of the dielectric resonators are significantly smaller than the dimensions of the metal resonators. Therefore, in recent years there has been a tendency to replace metal elements with sub-wave weakly absorbing dielectric structures from metamaterials, and the problem has arisen of controlling the magnetic components of the light field of the optical and infrared ranges using dielectric magnetic structures. Thus, the task is to create dielectric structures with the required spectral and spatial characteristics of the magnetic response in the gigahertz, terahertz and infrared ranges.

Direct experimental studies of the electromagnetic properties for individual nodes of nanoscale structures are difficult. Therefore, nanostructures with alternating magnetic susceptibility can be pre-modeled using theoretically calculated and experimentally verified parameters of the dielectric magnetic dipoles of the gigahertz range.

The purpose of the experiments was to directly measure the fields near the dielectric frame for correct comparison with the results of theoretical calculations. The main tasks that were set in the experiment: the excitation of a resonant electromagnetic response with the phase change in the reradiated wave and the inversion of the resulting field; detection of resonant frequencies when a linearly polarized TEM-wave is incident on the dielectric frame; investigation of the influence of material and object geometry on resonant frequencies.

The size of the frame was preliminarily chosen so that, on the one hand, it was large enough to measure the fields around the frame, and, on the other hand, so that it was much smaller than the wavelength of the incident wave for the region of the assumed resonant frequencies. In doing so, the studied frequency band should be accessible and convenient enough for measurements. As a result, the following frame characteristics were selected (see Fig. 3): external dimensions is 3 cm × 2 cm (therefore $a = 2.2$ cm, $b = 1.2$ cm, $a_1 = 2.6$ cm, $b_1 = 1.6$ cm, $a_2 = 3$ cm, $b_2 = 2$ cm), square cross-section is $h = 0.4$ cm, relative dielectric constant is $\varepsilon \approx 155$.

Gigahertz Resonant Measurement. Agilent E5071C ENA Network Analyzer was used for the generation and registration of emission spectra of GHz-range. In one case ETS-Lindgren's model 3160-09 pyramidal horn antennas were employed as the transmitter and receiver. In the other case ETS-Lindgren's model 3115 double ridged waveguide horn antenna was applied as a transmitter. To increase the signal-to-noise ratio and decrease the influence of radio noise the additional amplifier was used. The transmitter power was 10 mw. The linear probes of high frequency electric field with the length of sensitive element 1 cm were used for measurements of electric field near by the frame. The high frequency magnetic field in the near zone was registered by the screened circular probe with 0.5 cm diameter of sensitive element.

Phase Shift Measurement. The study of the phase shift between the initial and the reradiated waves was performed using a high-speed four-channel oscilloscope Tektronix DPO73304DX with two probes of the electric field. One probe located near antenna beyond of the area of influence of dielectric frame was used as a referent one. Its signal was recorded by one channel of the oscillations detector. Another probe was located close to the frame. Its signal was fixed by the second channel. Both probes were oriented in the direction of maximum sensitivity to the electric field of the incident wave and the distance between the probes kept constant in order to avoid any supplementary phase shift between them in the measurement process.

Extensive measurements were developed and carried out by a group of experimenters from the JIHT RAS (Moscow) composed of V.Ya. Pecherkin, L.M. Vasilyak, S.P. Vetchinin, and detailed measurement results will be published by them in a separate article (Comments for the second version on ArXiv.org: that article was published: A.B. Shvartsburg, V.Ya. Pecherkin, S. Jiménez, L.M. Vasilyak, L. Vázquez and S.P. Vetchinin, Resonant phenomena in an all-dielectric rectangular circuit induced by a plane microwave, J. Phys. D: Appl. Phys. **54**, 075004 (2021). DOI 10.1088/1361-6463/abc280). Our goal was to theoretically calculate resonances and compare them with experimental data. As a result, we can state that the measured experimental values of resonant frequencies are in good agreement with the theoretical values found in previous Sects.

## 5. Conclusion

Experimental and theoretical studies of electromagnetic low frequency *LC* resonances in all-dielectric rectangular frames excited by incident linearly polarized microwaves subject to the various orientations of these frames with respect to the polarization structure of the driving wave are carried out. The displacement currents induced by this wave in both whole frame and its parallel sides are shown to provide the formation of several magnetic and electric dipoles, characterized by different resonant frequencies. Consideration of these frames located in three orthogonal planes, normal respectively to electric component $\vec{E}$, magnetic component $\vec{H}$ and wave vector $\vec{k}$ of the driving wave (see Figs. 5, 3 and 4) had revealed the following salient features of resonant magnetic and electric dipoles:

1. The *LC* magnetic resonances of all-dielectric rectangular frame are interchanged by electric ones, forming the complicated spectrum of electromagnetic resonances. These resonances are habitual to the quasi stationary fields in the near zone of the scattering frame.

2. Subject to the orientation of rectangle frame with respect to vector trio $\vec{E}$, $\vec{H}$, $\vec{k}$ characterizing the driving wave, the induced magnetic and electric dipoles possess the different resonant frequencies: (6), (13), (18), (20) and (22). Resonant inversion of magnetic inductance and electric displacement stipulated by these dipoles occurs in the high frequency spectral ranges $f \geq f_0$.

3. The turn of rectangular frame from position A to position B (see Figs. 4 and 5), accompanied by the variation of resonant frequency, indicates the angular anisotropy and frequency dispersion of induced electric dipoles.

4. When the resonant oscillations are excited in the parallel sides of rectangular frame, spaced by the half length of driving wave, this circuit can be viewed as a peculiar structure - all-dielectric magnetic or electric resonant quadrupole.

The theory of these phenomena is elaborated from the first principles. Calculations of resonant frequencies are in good agreement with the experimental data. All-dielectric LC frames are designed and tested in the GHz range. Such frames visualizing the spectral properties of lonely oscillating element ensure the possibility to model the near fields of nanoscale LC circuits inaccessible for the direct measurements now. The practical usefulness of the obtained results is connected with extremely small losses in dielectric structures as compared with the metallic ones. Due to smallness of losses the Q-factor of all-dielectric structures is increased and the width of resonance is narrowed.

**Appendix: Formulas for capacitances and inductances**

Precise formula for inductance of a wire [21]:

$$L = N_l - G_g + A_a - Q_q, \quad G_g = 2l \ln r_g, \quad A_a = 2r_a, \quad Q_q = \frac{1}{2D} r_q^2, \qquad (A1)$$

where $r_g$ is the geometric mean distance of the cross-sectional area from itself, $r_a$ is the arithmetic mean distance of the cross-sectional area from itself, and $r_q$ is the mean square distance of the cross-sectional area from itself, $D$ is the distance between the ends of the wire; for a straight wire $D = l$, $N_l = 2l(\ln 2l - 1)$.

Self-induction of a rectilinear wire with length $l$ and square section $h \times h$ is:

$$L = 2l \left( \ln \frac{l}{h} + \frac{1}{2} \right). \qquad (A2)$$

Mutual induction of parallel wires of length $l$ spaced $d$ is:

$$M = 2l \left( \ln \frac{l + \sqrt{l^2 + d^2}}{d} - \frac{\sqrt{l^2 + d^2}}{l} + \frac{d}{l} \right). \qquad (A3)$$

When a unidirectional current is excited, the self-induction coefficient $L$ of a rectangular frame can be found according to the general rules [21], taking into account the self-induction of two sides of the rectangular frame with the length $a_1$ and two sides of this frame with the length $b_1$, as well as mutual induction in each pair (taking into account the opposite direction of currents on opposite sides framework). Denoting $p = b_1 / a_1$, final expression for the self-inductance of rectangular frame is $L = 4a_1 K$, where

$$K = \ln \frac{a_1}{h} - \frac{1}{2} - \ln \frac{1 + \sqrt{1 + p^2}}{p} + 2\sqrt{1 + p^2} + p \left( \ln \frac{b_1}{h} - \frac{1}{2} - \ln \left[ p + \sqrt{1 + p^2} \right] \right). \qquad (A4)$$

The capacitance of flat capacitor is

$$C = \frac{\varepsilon S}{4\pi d}, \qquad (A5)$$

where $\varepsilon$ is the dielectric permittivity, $S$ is the area, $d$ is the distance between the plates.

In our case, the capacitance $c_a$ per unit length (as two pair of plates with $S = h^2$, $d = a_2$) is:

$$c_a = \frac{\varepsilon h^2}{2\pi a_2^2}. \qquad (A6)$$

The self-inductance $l_a$ per unit length for two wires is:

$$l_a = 4 \ln \frac{a_2}{h} + 2. \qquad (A7)$$

For two wires with parallel and anti-parallel currents respectively, the mutual inductance $\pm m_a$ per unit length is:

$$\pm m_a = \pm 2\left( \ln \frac{a_2 + \sqrt{a_2^2 + b_1^2}}{b_1} - \frac{\sqrt{a_2^2 + b_1^2}}{a_2} + \frac{b_1}{a_2} \right). \tag{A8}$$